\documentclass[aps,prb,amsmath,amssymb]{revtex4}
\usepackage{graphicx}% Include figure files
\usepackage{bm}% bold math
\usepackage{bbold}
\usepackage{hhline}

\def\br{ \bm{r} }
\def\bk{ \bm{k} }
\def\bmd{ \bm{d} }
\def\be{ \bm{e} }
\def\hbx{ \hat{\bm{x}} }
\def\hby{ \hat{\bm{y}} }
\def\hbz{ \hat{\bm{z}} }

\begin{document}
\title{Symmetry of superconducting pairing in non-pseudospin electron bands}

\author{K. V. Samokhin\footnote{E-mail: kirill.samokhin@brocku.ca}}
\affiliation{Department of Physics, Brock University, St. Catharines, Ontario L2S 3A1, Canada}

%\date{\today}

\begin{abstract}
We develop the symmetry classification of superconducting gap functions in electron bands that do not transform under the crystal point group operations like the pure spin-$1/2$ states. 
The Bloch state bases in twofold degenerate bands with spin-orbit coupling are defined across the Brillouin zone in the way which satisfies the symmetry and continuity requirements. These bases are used to 
construct general multiband pairing Hamiltonians in centrosymmetric crystals. Focusing on single-band pairing, four exceptional cases are identified in which the triplet gap function does not transform 
under the point group operations as a pseudovector, with a significant impact on the nodal structure.
\end{abstract}

\maketitle

\section{Introduction}
\label{sec: Intro}

The symmetry-based approach to the phenomenology of unconventional pairing states is a very powerful tool in the studies of fermionic superfuilds and superconductors.\cite{SU-review,TheBook} Using this approach, 
one can determine possible stable states and gap structures even if the microscopic details, in particular, the pairing mechanism, are not reliably known. At the heart of the symmetry classification of 
crystalline superconductors is the idea that the electron Bloch states transform under the point group operations and time reversal (TR) in the same way as the pure spin-$1/2$ states, 
even in the presence of spin-orbit coupling (SOC), see Refs. \onlinecite{And84,VG85,UR85}. 

The argument begins with the observation\cite{Kittel-book} that the electron bands in a TR invariant centrosymmetric crystal are twofold degenerate at each wave vector $\bk$ in the 
first Brillouin zone (BZ), because the Bloch states $|\bk\rangle$ and $KI|\bk\rangle$ belong to the same $\bk$ and are orthogonal ($K$ is the TR operation acting on spinors and $I$ is inversion).
Assuming that these states can be obtained from the pure spin-$1/2$ states $|\uparrow\rangle$ and $|\downarrow\rangle$ by turning on the SOC adiabatically, one can label them by the ``pseudospin'' index $s=1,2$. 
Quasiparticles in the four degenerate states $|\bk\rangle$, $KI|\bk\rangle$, $K|\bk\rangle$, and $I|\bk\rangle$ pair up to form either pseudospin-singlet or pseudospin-triplet Cooper pairs. 
To obtain the point-group symmetry properties of the pair wave functions, one has to
define the mutual orientations of the Bloch bases at different wave vectors corresponding to the rays of the star of $\bk$. The commonly used convention is given by the Ueda-Rice formula,\cite{UR85} in which the pseudospin bases
at the wave vectors $\bk$ and $g\bk$, where the point group operation $g$ is either a proper rotation $R$ or an improper rotation $IR$, 
are related by the same spin rotation matrix $\hat D^{(1/2)}(R)$ as the one that describes the transformation of the pure spin-$1/2$ states 
(recent discussions of the ways to construct the pseudospin bases across the BZ can be found in Refs. \onlinecite{Yip13} and \onlinecite{Fu15}). 
As a result, the triplet gap function $\bmd(\bk)$ transforms under the point group as a pseudovector, with well studied ramifications for the gap structure and other observable quantities.\cite{SU-review,TheBook} 

The approach outlined above hinges on the assumption that the Bloch states $|\bk\rangle$ and $KI|\bk\rangle$ transform like $|\uparrow\rangle$ and $|\downarrow\rangle$, which has to be true, in particular, at the center of the BZ
(the $\Gamma$ point). The symmetry group at the $\Gamma$ point is given by the point group $\mathbb{G}$ of the crystal and the states $|\bm{0}\rangle$ and $KI|\bm{0}\rangle$ form the basis of a two-dimensional (2D) 
double-valued representation of $\mathbb{G}$ (the fact that $K$ and $KI$ are antiunitary operations introduces some important complications, see Sec. \ref{sec: Bloch basis} below, but those can be ignored at the moment). Using 
the group character tables, see, \textit{e.g.}, Refs. \onlinecite{Lax-book} and \onlinecite{BC-book}, it is straightforward to check that not all such representations are equivalent to the spin-$1/2$ representation.
If the electron states at the $\Gamma$ point do not transform like the pure spin states, then the Ueda-Rice formula cannot be used to 
construct the Bloch bases continuously across the BZ, which is certain to have profound consequences for the symmetry classification of superconductors. Note that the effects of non-pseudospin character 
of the bands can be seen already in the normal state, for example, in the form of the antisymmetric (Rashba) SOC in crystals without an inversion center.\cite{Smidman-review,Sam19} 

One class of systems outside the scope of the pseudospin-based approach comprises the materials in which the Bloch bands are fourfold degenerate at the $\Gamma$ point, corresponding to a four-dimensional (4D) 
double-valued representation of one of the cubic point groups. Such band structure is realized, \textit{e.g.}, in the half-Heusler compounds YPtBi and LuPtBi, 
for which the ``$j=3/2$'' pairing has been suggested.\cite{j-3-2-pairing} Various issues with the standard classification of superconducting states in multiorbital 
systems have been recently discussed,\cite{WW09,Fis13,RS16,NHI16} especially in the context of the iron-based superconductors and Sr$_2$RuO$_4$.
Also, it has been shown that Blount's theorem\cite{Blount85} about the absence of line gap nodes in triplet superconductors does not hold in nonsymmorphic crystals, in which    
symmetry-protected line nodes can appear on the BZ boundary.\cite{non-symm}

Our goal is to extend the symmetry classification of superconducting states in twofold degenerate bands in centrosymmetric symmorphic crystals to the cases when the pseudospin description of the electron Bloch states 
is not applicable. The article is organized as follows. In Sec. \ref{sec: Bloch basis}, we discuss the reasons for the failure of the pseudospin-$1/2$ picture and show how the Ueda-Rice prescription should be modified in
non-pseudospin bands. 
In Sec. \ref{sec: SC symmetry}, the effects of the non-pseudospin character of the bands on superconducting pairing are discussed in the general multiband case. In Sec. \ref{sec: one band}, we focus on the single-band 
case and examine in detail the exceptional cases in which the standard symmetry classification of triplet states breaks down.
Throughout the paper we use the units in which $\hbar=1$.

\section{Bloch basis in non-pseudospin bands}
\label{sec: Bloch basis}

Let us consider a TR invariant centrosymmetric crystal described by a symmorphic space group (the last condition can be relaxed if one focuses on the momentum dependence of the superconducting gap function in the BZ interior).  
The electron Bloch states are at least twofold degenerate at each $\bk$ due to the combined symmetry operation ${\cal C}=KI$, 
called conjugation.\cite{Kittel-book} Namely, the states $|\bk,n,1\rangle$ and $|\bk,n,2\rangle\equiv{\cal C}|\bk,n,1\rangle$ are orthogonal and have the same energy. 
The index $n$ labels the bands, while the additional index $s=1,2$ distinguishes two orthonormal states within the same band. Due to the inevitable presence of the electron-lattice SOC, 
the conjugacy index $s$ does not correspond to the electron spin projection. 

Since the Bloch states $|\bk,n,s\rangle$ are spin-$1/2$ wave functions, we have ${\cal C}^2=-1$ and ${\cal C}|\bk,n,2\rangle=-|\bk,n,1\rangle$, which is the same as the action of conjugation on 
the pure spin eigenstates $|\uparrow\rangle\equiv\xi_1$ and $|\downarrow\rangle\equiv\xi_2$. However, the analogy between $|\bk,n,1\rangle,|\bk,n,2\rangle$ and $\xi_1,\xi_2$ does not always extend to
the transformation under the point group rotations and reflections. From the group-theoretical point of view, the states $|\bk,n,1\rangle$ and $|\bk,n,2\rangle$ form the 
basis of an irreducible double-valued corepresentation (corep) of the magnetic point group of the wave vector $\bk$, which is not always equivalent to the spin-$1/2$ corep.
The full symmetry group of $\bk$ is ``magnetic'', because it contains the antiunitary operation ${\cal C}$ (a detailed review of magnetic groups and their coreps can be found, \textit{e.g.}, in Refs. \onlinecite{BC-book} 
and \onlinecite{BD68}). 

At the $\Gamma$ point, the magnetic symmetry group is ${\cal G}=\mathbb{G}+{\cal C}\mathbb{G}$. Since we consider only crystals with a center of inversion, the point group $\mathbb{G}$ can be represented as a direct 
product of some other (noncentrosymmetric) point group $\tilde{\mathbb{G}}$ and $\mathbf{C}_i=\{E,I\}$. Therefore, the coreps of ${\cal G}$ are either inversion-even ($\Gamma^+$) or inversion-odd ($\Gamma^-$). 
Clearly, only even coreps can be equivalent to the spin-$1/2$ corep. 
The double-valued coreps of all centrosymmetric magnetic point groups are listed in Table \ref{table: Gamma-point-coreps}, using the standard notations for the irreducible representations (irreps), see, \textit{e.g.}, 
Refs. \onlinecite{Lax-book} and \onlinecite{BC-book}. 
Note that all these coreps are 2D, except $(\Gamma_6^\pm,\Gamma_7^\pm)$ for $\mathbb{G}=\mathbf{T}_{h}$ and $\Gamma_8^\pm$ for $\mathbb{G}=\mathbf{O}_{h}$, 
which are 4D. The 4D coreps correspond to the bands which are fourfold degenerate at the $\Gamma$ point, for instance, the $\Gamma_8^\pm$ (``$j=3/2$'') bands for $\mathbb{G}=\mathbf{O}_{h}$ (Ref. \onlinecite{Lutt56}), that 
will not be considered here. Pairs of complex conjugate irreps $(\Gamma,\Gamma^*)$ produce coreps of the ``pairing'' type of twice the dimension, while the one-dimensional (1D) irreps $\Gamma_2^\pm$ of $\mathbb{G}=\mathbf{C}_{i}$
and $\Gamma_6^\pm$ of $\mathbb{G}=\mathbf{C}_{3i}$ produce 2D coreps of the ``doubling'' type.\cite{Lax-book}

\begin{table}
\caption{The double-valued coreps of the centrosymmetric point groups at the $\Gamma$ point. The last column shows whether the inversion-even corep ($\Gamma^+$) is equivalent to the spin-1/2 corep.}
\begin{tabular}{|c|c|c|c|}
    \hline
    $\quad \mathbb{G}\quad $ & $\quad \tilde{\mathbb{G}}\quad $ & corep & \ pseudospin\ \ \\ \hline
    $\mathbf{C}_{i}$  & $\mathbf{C}_{1}$ & $\Gamma_2$ &  Y  \\ \hline
    $\mathbf{C}_{2h}$  & $\mathbf{C}_{2}$ & $(\Gamma_3,\Gamma_4)$ &  Y \\ \hline
    $\mathbf{D}_{2h}$  & $\mathbf{D}_{2}$ & $\Gamma_5$   &  Y  \\ \hline
    $\mathbf{C}_{4h}$  & $\mathbf{C}_{4}$ & $(\Gamma_5,\Gamma_6)$  &  Y \\ 
                       & & $(\Gamma_7,\Gamma_8)$  &  N  \\ \hline
    $\mathbf{D}_{4h}$  & $\mathbf{D}_{4}$ & $\Gamma_6$  &  Y \\ 
		       & & $\Gamma_7$  &   N \\ \hline
    $\mathbf{C}_{3i}$  & $\mathbf{C}_{3}$ & $(\Gamma_4,\Gamma_5)$  & Y  \\ 
                       & & $\Gamma_6$  & N  \\ \hline		       
    $\mathbf{D}_{3d}$  & $\mathbf{D}_{3}$ & $\Gamma_4$  &  Y  \\ 
		      & & $(\Gamma_5,\Gamma_6)$  & N  \\ \hline
    $\mathbf{C}_{6h}$  & $\mathbf{C}_{6}$ & $(\Gamma_7,\Gamma_8)$  &  Y  \\ 
		      & & $(\Gamma_9,\Gamma_{10})$  & N  \\
                       & & $(\Gamma_{11},\Gamma_{12})$ &   N  \\ \hline			
    $\mathbf{D}_{6h}$  & $\mathbf{D}_{6}$ & $\Gamma_7$ &   Y  \\ 
                       & & $\Gamma_8$ &  N  \\ 
                       & & $\Gamma_9$ &  N  \\ \hline
    $\mathbf{T}_{h}$  & $\mathbf{T}$ & $\Gamma_5$ &  Y  \\
		       & & $(\Gamma_6,\Gamma_7)$ &   N  \\ \hline                       
    $\mathbf{O}_{h}$  & $\mathbf{O}$ & $\Gamma_6$ &   Y  \\
		       & & $\Gamma_7$ &  N  \\ 
                       & & $\Gamma_8$ &  N  \\ \hline		
\end{tabular}
\label{table: Gamma-point-coreps}
\end{table}

We assume that the Bloch states at the $\Gamma$ point in the $n$th band, $|\bm{0},n,1\rangle$ and $|\bm{0},n,2\rangle$, form the basis of a 2D double-valued corep of ${\cal G}$ described by 
$2\times 2$ matrices $\hat{\cal D}_n(g)$, where $g\in\mathbb{G}$. Then, one can construct the Bloch bases at $\bk\neq\bm{0}$ using the following expression:\cite{Sam19}
\begin{equation}
\label{general-prescription}
    g|\bk,n,s\rangle=\sum_{s'}|g\bk,n,s'\rangle {\cal D}_{n,s's}(g),
\end{equation}
which defines the basis at $g\bk$ given the basis at $\bk$. In particular, 
\begin{equation}
\label{general-prescription-I}
    I|\bk,n,s\rangle=p_n|-\bk,n,s\rangle,
\end{equation}
where $p_n=\pm$ denotes the parity of the corep. Therefore, we have
\begin{equation}
\label{general-prescription-K}
    K|\bk,n,1\rangle=p_n|-\bk,2\rangle,\quad K|\bk,2\rangle=-p_n|-\bk,n,1\rangle
\end{equation}
for the transformation under TR operation $K={\cal C}I$. From Eq. (\ref{general-prescription}) we obtain the general point-group transformation rules for the electron creation operators in the Bloch states in the $n$th band:
\begin{equation}
\label{c-dagger-transform}
  gc^\dagger_{\bk,n,s}g^{-1}=\sum_{s'}c^\dagger_{g\bk,n,s'}{\cal D}_{n,s's}(g),
\end{equation}
for all $g\in\mathbb{G}$.

The prescription (\ref{general-prescription}) can be justified as follows. Let us pick a wave vector $\bk$ in the fundamental domain of the BZ and apply the point group element $g$ to transform $\bk$ into $g\bk$ 
-- a ray of the star of $\bk$. Since the state $g|\bk,n,s\rangle$ belongs to the wave vector $g\bk$, it can be represented in the form
\begin{equation}
\label{basis-transform-U}
  g|\bk,n,s\rangle=\sum_{s'}|g\bk,n,s'\rangle U_{n,s's}(\bk;g),
\end{equation}
where the expansion coefficients form a unitary matrix. If one assumes that this matrix can be chosen to be $\bk$ independent, then its form is fixed by putting $\bk=\bm{0}$ in Eq. (\ref{basis-transform-U}):
$$
  g|\bm{0},n,s\rangle=\sum_{s'}|\bm{0},n,s'\rangle U_{n,s's}(g),
$$
which yields $\hat U_n(g)=\hat{\cal D}_n(g)$.

If the $\Gamma$-point corep is equivalent to the spin-$1/2$ corep, see the last column of Table \ref{table: Gamma-point-coreps}, then the band is called ``pseudospin band'' and one can put
\begin{equation}
\label{Ueda-Rice}
  \hat{\cal D}_n(g)=\hat D^{(1/2)}(R)
\end{equation}
for $g=R$ or $IR$, where $\hat D^{(1/2)}(R)=e^{-i\theta(\bm{n}\hat{\bm{\sigma}})/2}$ is the spin-1/2 representation of a counterclockwise rotation $R$ through an angle $\theta$ about an axis $\bm{n}$ and 
$\hat{\bm{\sigma}}=(\hat\sigma_1,\hat\sigma_2,\hat\sigma_3)$ are the Pauli matrices. In particular,
$I|\bk,n,s\rangle=|-\bk,n,s\rangle$. The expression (\ref{Ueda-Rice}) constitutes the Ueda-Rice convention for the pseudospin Bloch basis.\cite{UR85}
In general, however, the Bloch states at the $\Gamma$ point correspond to a corep which is not equivalent to the spin-$1/2$ corep and the Ueda-Rice prescription does not work. In this case, 
the band can be called ``non-pseudospin band''. 

The corep matrices for the 2D non-pseudospin coreps at the $\Gamma$ point are shown in Table \ref{table: non-pseudospin-corep-matrices}, see also Appendix \ref{app: D_3d-coreps}. 
For the uniaxial point groups, the corep basis functions are chosen to be eigenfunctions of the total angular momentum $\hat j_z$.  
Since each element of $\mathbb{G}=\tilde{\mathbb{G}}\times\mathbf{C}_i$ has the form $g=\tilde g$ or $g=I\tilde g$, where $\tilde g\in\tilde{\mathbb{G}}$, 
one can use the expressions $\hat{\cal D}_{\Gamma^\pm}(\tilde g)=\hat{\cal D}_{\Gamma}(\tilde g)$ and $\hat{\cal D}_{\Gamma^\pm}(I\tilde g)=\pm\hat{\cal D}_{\Gamma}(\tilde g)$ and 
Table \ref{table: non-pseudospin-corep-matrices} to obtain the corep matrices for all elements of the point group.

\begin{table}
\caption{The 2D non-pseudospin double-valued coreps at the $\Gamma$ point, with examples of even and odd conjugate basis functions corresponding to the lowest possible value of the total angular momentum $j_z$ 
(the last column); $\tilde g$ denotes the generators of $\tilde{\mathbb{G}}$ and $\rho_\pm=x\pm iy$. The spin quantization axis is chosen along $\hbz$. 
In the last row, $f(\br)$ is a real basis function of the $\Gamma^+_2$ or $\Gamma^-_2$ irrep of $\mathbf{O}_h$ (changing sign under a $C_{4z}$ rotation), 
\textit{e.g.}, $f_{\Gamma_2^+}=x^4(y^2-z^2)+y^4(z^2-x^2)+z^4(x^2-y^2)$ and $f_{\Gamma_2^-}=xyz$.}
\begin{tabular}{|c|c|c|c|c|c|}
    \hline
    $\quad \mathbb{G}\quad $ &  corep $\Gamma$ & $\hat{\cal D}_\Gamma(\tilde g)$ & even basis & odd basis & $j_z$ \\ \hline
    $\mathbf{C}_{4h}$  & $(\Gamma_7,\Gamma_8)$   &  $\hat{\cal D}(C_{4z})=-\hat D^{(1/2)}(C_{4z})$ & $\rho_-^2\xi_1$, $\rho_+^2\xi_2$ & $\rho_+\xi_1$, $-\rho_-\xi_2$  &\ $\pm 3/2$\ \ \\ \hline
    $\mathbf{D}_{4h}$ & $\Gamma_7$  & \ $\hat{\cal D}(C_{4z})=-\hat D^{(1/2)}(C_{4z})$,\quad $\hat{\cal D}(C_{2y})=\hat D^{(1/2)}(C_{2y})$\ \ &  $\rho_-^2\xi_1$, $\rho_+^2\xi_2$ & $\rho_+\xi_1$, $-\rho_-\xi_2$ & $\pm 3/2$  \\ \hline
    $\mathbf{C}_{3i}$  & $\Gamma_6$ &  $\hat{\cal D}(C_{3z})=-\hat\sigma_0$ & $\rho_-^2\xi_1$, $\rho_+^2\xi_2$ &  $\rho_+\xi_1$, $-\rho_-\xi_2$ & $\pm 3/2$ \\ \hline
    $\mathbf{D}_{3d}$  & $(\Gamma_5,\Gamma_6)$  & $\hat{\cal D}(C_{3z})=-\hat\sigma_0$,\quad $\hat{\cal D}(C_{2y})=\hat D^{(1/2)}(C_{2y})$ & $\rho_-^2\xi_1$, $\rho_+^2\xi_2$ &  $\rho_+\xi_1$, $-\rho_-\xi_2$ & $\pm 3/2$ \\ \hline								
    $\mathbf{C}_{6h}$  & $(\Gamma_{9},\Gamma_{10})$  & $\hat{\cal D}(C_{6z})=-\hat D^{(1/2)}(C_{6z})$ & $\rho_+^2\xi_1$, $\rho_-^2\xi_2$  & $\rho_-^3\xi_1$, $-\rho_+^3\xi_2$ & $\pm 5/2$ \\ \cline{2-6}
                      & $(\Gamma_{11},\Gamma_{12})$  & $\hat{\cal D}(C_{6z})=-\hat D^{(1/2)}(C_{2z})$ &  $\rho_-^2\xi_1$, $\rho_+^2\xi_2$ &  $\rho_+\xi_1$, $-\rho_-\xi_2$ & $\pm 3/2$ \\ \hline
    $\mathbf{D}_{6h}$  & $\Gamma_8$  & $\hat{\cal D}(C_{6z})=-\hat D^{(1/2)}(C_{6z})$,\quad $\hat{\cal D}(C_{2y})=\hat D^{(1/2)}(C_{2y})$ & $\rho_+^2\xi_1$, $\rho_-^2\xi_2$ & $\rho_-^3\xi_1$, $-\rho_+^3\xi_2$ & $\pm 5/2$ \\ \cline{2-6}
                       & $\Gamma_9$ & $\hat{\cal D}(C_{6z})=-\hat D^{(1/2)}(C_{2z})$,\quad $\hat{\cal D}(C_{2y})=\hat D^{(1/2)}(C_{2y})$ &  $\rho_-^2\xi_1$, $\rho_+^2\xi_2$ &  $\rho_+\xi_1$, $-\rho_-\xi_2$ & $\pm 3/2$ \\ \hline
    $\mathbf{O}_h$  & $\Gamma_7$ &  $\hat{\cal D}(C_{4z})=-\hat D^{(1/2)}(C_{4z})$,\quad $\hat{\cal D}(C_{2y})=\hat D^{(1/2)}(C_{2y})$ & \ $f_{\Gamma_2^+}(\br)(\xi_1,\xi_2)$\ \  &\ $if_{\Gamma_2^-}(\br)(\xi_1,\xi_2)$\ \ &  \\ 
                                     & &  $\hat{\cal D}(C_{3xyz})=\hat D^{(1/2)}(C_{3xyz})$ & & &  \\ \hline                                                               
\end{tabular}
\label{table: non-pseudospin-corep-matrices}
\end{table}

\section{Pairing symmetry: General analysis}
\label{sec: SC symmetry}

In this section, the Bloch bases defined by Eq. (\ref{general-prescription}) are used to analyze the symmetry of superconducting pairing.
Suppose there are $N$ bands crossing the chemical potential, then the pairing can be described by the following general mean-field Hamiltonian:
\begin{equation}
\label{H-mean-field-general}
  \hat H=\sum_{\bk,n,s}\epsilon_n(\bk)c^\dagger_{\bk,n,s}c_{\bk,n,s}+\frac{1}{2}\sum_{\bk,nn',ss'}\left[\Delta_{nn',ss'}(\bk)c^\dagger_{\bk,n,s}\tilde c^\dagger_{\bk,n',s'}+\mathrm{H.c.}\right],
\end{equation}
where $n=1,...,N$ and $\epsilon_n(\bk)=\epsilon_n(-\bk)$ are the band dispersions. The second term, which we denote below as $\hat H_{sc}$, describes the pairing between time-reversed states,\cite{And59,Blount85} 
with
\begin{equation}
\label{tilde-c}
  \tilde c^\dagger_{\bk,n,s}=Kc^\dagger_{\bk,n,s}K^{-1}=p_n\sum_{s'}c^\dagger_{-\bk,n,s'}(-i\hat\sigma_2)_{s's}, 
\end{equation}
according to Eq. (\ref{general-prescription-K}). At each $\bk$, the gap functions $\hat\Delta(\bk)$ are $2N\times 2N$ 
matrices in the band and conjugacy spaces. The matrix elements $\hat\Delta_{nn}$ describe the intraband pairing in the $n$th band, while $\hat\Delta_{nn'}$ with $n\neq n'$ describe the interband pairing.

The matrix structure of the gap functions in the conjugacy space can be represented in the form
\begin{equation}
\label{singlet-triplet}
  \hat\Delta_{nn'}(\bk)=\psi_{nn'}(\bk)\hat\sigma_0+\bmd_{nn'}(\bk)\hat{\bm{\sigma}},
\end{equation}
where $\hat\sigma_0$ is the $2\times 2$ unit matrix.
By analogy with the standard theory of unconventional superconductivity,\cite{SU-review,TheBook} this last expression can be called the singlet-triplet decomposition (one should keep in mind that $s$ is neither spin nor 
pseudospin, in general). The parity of the singlet and triplet components can be obtained by substituting Eq. (\ref{tilde-c}) in the Hamiltonian (\ref{H-mean-field-general}) and using the anticommutation of the fermionic 
operators, with the following result:
\begin{equation}
\label{parity-general}
  \psi_{nn'}(-\bk)=p_np_{n'}\psi_{n'n}(\bk),\quad \bmd_{nn'}(-\bk)=-p_np_{n'}\bmd_{n'n}(\bk).
\end{equation}
Thus the intraband singlet (triplet) pairing is always even (odd) in $\bk$, while the interband components do not have a definite parity. 

Regarding the symmetry under the point-group rotations and reflections $g$, we use Eq. (\ref{c-dagger-transform}) as well the antilinearity of the TR operator and the fact that it commutes with all $g$ to obtain:
$$
  g\tilde c^\dagger_{\bk,n,s}g^{-1}=g(Kc^\dagger_{\bk,n,s}K^{-1})g^{-1}=K(gc^\dagger_{\bk,n,s}g^{-1})K^{-1}=\sum_{s'}\tilde c^\dagger_{g\bk,n,s'}{\cal D}^*_{n,s's}(g).
$$
Substituting this in $g\hat H_{sc}g^{-1}$, we arrive at the transformation rules for the gap functions:
\begin{equation}
\label{Delta-transform-general-g}
  g:\ \hat\Delta_{nn'}(\bk)\to\hat{\cal D}_n(g)\hat\Delta_{nn'}(g^{-1}\bk)\hat{\cal D}^\dagger_{n'}(g).
\end{equation}
In terms of the singlet and triplet components, this becomes
\begin{equation}
\label{singlet-triplet-transform-g}
  g:\ \psi_{nn'}(\bk)\hat\sigma_0+\bmd_{nn'}(\bk)\hat{\bm{\sigma}}\to \psi_{nn'}(g^{-1}\bk)\left[\hat{\cal D}_n(g)\hat{\cal D}^\dagger_{n'}(g)\right]
    +\bmd_{nn'}(g^{-1}\bk)\left[\hat{\cal D}_n(g)\hat{\bm{\sigma}}\hat{\cal D}^\dagger_{n'}(g)\right].
\end{equation}
We see that, while the intraband ($n=n'$) singlet and triplet gap functions transform independently of each other, it might not be the case for the interband gaps.
Finally, under the TR operation, using $K\tilde c^\dagger_{\bk,n,s}K^{-1}=-c^\dagger_{\bk,n,s}$ and the anticommutation of the fermionic operators, we obtain:
$$
  K\hat H_{sc}K^{-1}=\frac{1}{2}\sum_{\bk,nn',ss'}\Delta_{n'n,s's}^*(\bk)c^\dagger_{\bk,n,s}\tilde c^\dagger_{\bk,n',s'}+\mathrm{H.c.}
$$
This means that $\Delta_{nn',ss'}(\bk)$ is transformed into $\Delta_{n'n,s's}^*(\bk)$, therefore,
\begin{equation}
\label{singlet-triplet-transform-K}
  K:\quad \psi_{nn'}(\bk)\to\psi_{n'n}^*(\bk),\quad \bmd_{nn'}(\bk)\to\bmd_{n'n}^*(\bk). 
\end{equation}
It follows from Eq. (\ref{singlet-triplet-transform-g}) that in general the singlet and triplet gap functions do not transform as scalars and 
vectors under the point group operations. To illustrate the effects of the non-pseudospin character of the bands on superconducting pairing, in the next section we focus on the simplest case of just one twofold degenerate band.

\section{Single band pairing}
\label{sec: one band}

One can drop the band indices in Eqs. (\ref{parity-general}), (\ref{singlet-triplet-transform-g}), and (\ref{singlet-triplet-transform-K}), and obtain that the singlet component $\psi$ is even in $\bk$ 
and the triplet component $\bmd$ is odd, while the action of TR on $\psi$ and $\bmd$ is equivalent to complex conjugation. Under the point group operations, the gap functions transform as follows:
\begin{equation}
\label{singlet-triplet-g-one-band}
  g:\ \psi(\bk)\to\psi(g^{-1}\bk),\quad \bmd(\bk)\to{\cal R}(g)\bmd(g^{-1}\bk),
\end{equation}
where the $3\times 3$ orthogonal matrix $\hat{\cal R}$ is defined by
\begin{equation}
\label{cal R-def}
  \hat{\cal D}_\Gamma^\dagger(g)\hat{\sigma}_i\hat{\cal D}_\Gamma(g)=\sum_{j=1}^3{\cal R}_{ij}(g)\hat{\sigma}_j.
\end{equation}
Note that $\hat{\cal R}(g)$ does not depend on the parity of the band.

It follows from Eq. (\ref{singlet-triplet-g-one-band}) that $\psi$ always transforms under the point group as a complex scalar, regardless of the band symmetry at the $\Gamma$ point. Therefore, the usual classification of
the singlet superconducting states\cite{SU-review,TheBook} is applicable. The gap function can be represented in the following form: 
\begin{equation}
\label{psi-one-band-expansion}
  \psi(\bk)=\sum_{a=1}^{d}\eta_a\phi_a(\bk),
\end{equation}
where $a$ labels the scalar basis functions $\phi$ of a $d$-dimensional even single-valued irrep $\gamma$ of the point group $\mathbb{G}$. The expansion coefficients $\eta_a$, which play the role of the superconducting order 
parameter components, can be found by minimizing the free energy of the superconductor.

In contrast, the transformation properties of the triplet gap function essentially depend on the band symmetry at the $\Gamma$ point. 
In a pseudospin band, we have $\hat{\cal D}_\Gamma(g)=\hat{D}^{(1/2)}(R)$ for both proper ($g=R$) and improper ($g=IR$) rotations. Using the identity
\begin{equation}
\label{D-to-R}
  \hat{D}^{(1/2),\dagger}(R)\hat{\sigma}_i\hat{D}^{(1/2)}(R)=\sum_{j=1}^3R_{ij}\hat{\sigma}_j,
\end{equation}
where $\hat R\equiv\hat D^{(1)}(R)$ is the spin-$1$ rotation matrix, we obtain from Eq. (\ref{singlet-triplet-g-one-band}):
\begin{equation}
\label{d-transform-pseudospin}
  g:\ \bmd(\bk)\to R\bmd(g^{-1}\bk).
\end{equation}
Therefore, the triplet gap function transforms like a pseudovector and can be represented as a linear combination
\begin{equation}
\label{d-one-band-expansion}
  \bmd(\bk)=\sum_{a=1}^{d}\eta_a\bm{\phi}_a(\bk)
\end{equation}
of the ``spin-vector'' basis functions of an odd single-valued irrep $\gamma$ of $\mathbb{G}$. Similarly to the singlet case, the components of the superconducting order parameter $\eta_1,..,\eta_d$ 
are found by minimizing the free energy of the superconductor.
The expressions (\ref{d-transform-pseudospin}) and (\ref{d-one-band-expansion}) form the foundation of the standard symmetry-based treatment of triplet superconductivity.\cite{SU-review,TheBook} 
In particular, $\bmd(\bk)$ determines the gap in the Bogoliubov excitation spectrum: If $\bm{d}^2(\bk)=0$ along a line (or in a plane) intersecting the Fermi surface, 
then there is a point (or line) node in the gap. The latter possibility, \textit{i.e.}, a line node, cannot happen generically.\cite{Blount85} 

Let us recall how the spin-vector basis functions in Eq. (\ref{d-one-band-expansion}) are constructed.\cite{UR85,VG85} We introduce three unit vectors $\be_1\parallel\hbx$, 
$\be_2\parallel\hby$, and $\be_3\parallel\hbz$, which are not affected by inversion (\textit{i.e.}, they are pseudovectors) and transform under rotations as follows: $\be_i\to R\be_i=\sum_j\be_jR_{ji}$. 
These vectors span a three-dimensional representation $\tilde\gamma^+$, not necessarily irreducible, of $\mathbb{G}$. Next, we take a $r$-dimensional odd irrep $\gamma^-$, 
with the scalar basis functions $\varphi_1(\bk),...,\varphi_r(\bk)$. The $3r$ functions $\be_i\varphi_\mu(\bk)$ 
($i=1,2,3$; $\mu=1,...,r$) form the basis of the product representation $\tilde\gamma^+\times\gamma^-$, which can be decomposed into a sum of irreps of $\mathbb{G}$. For each irrep in this decomposition, 
the basis functions $\bm{\phi}$ are given by certain linear combinations of $\be_i\varphi_\mu(\bk)$. Going through all $\gamma^-$, one obtains all contributions to $\bm{\phi}_a(\bk)$.

One can see why this approach might fail in a non-pseudospin band: if the $\Gamma$-point corep is such that $\hat{\cal R}(g)\neq\hat R$, then the transformation of the triplet gap function under 
$g$ does not have the form (\ref{d-transform-pseudospin}). Using Table \ref{table: non-pseudospin-corep-matrices}, we obtain that this happens in the following four cases: 
\begin{equation}
\label{exceptional-coreps}
  \Gamma_6\ \mathrm{of}\ \mathbf{C}_{3i},\quad (\Gamma_5,\Gamma_6)\ \mathrm{of}\ \mathbf{D}_{3d},\quad 
  (\Gamma_{11},\Gamma_{12})\ \mathrm{of}\ \mathbf{C}_{6h},\quad \Gamma_9\ \mathrm{of}\ \mathbf{D}_{6h}.
\end{equation}
Thus the classification of the triplet superconducting states based on Eq. (\ref{d-transform-pseudospin}) is not applicable in certain bands in trigonal and hexagonal crystals.
This conclusion agrees with the observation made recently in Ref. \onlinecite{SY18} that there can be gap nodes not explained by the standard pseudospin-based approach 
on a threefold or sixfold axis in the BZ.

Below we present a detailed derivation of the momentum dependence of the triplet gap functions in the exceptional cases listed in Eq. (\ref{exceptional-coreps}). The results are summarized in Tables 
\ref{table: d-vector-C_3i}, \ref{table: d-vector-D_3d}, \ref{table: d-vector-C_6h}, and \ref{table: d-vector-D_6h}, which also include, for comparison, the corresponding expressions in the pseudospin bands. In all cases, 
the lowest-order polynomial expressions for each component of $\bmd(\bk)$ are shown, which are applicable in the vicinity of the $\Gamma$ point.

\subsection{$\mathbb{G}=\mathbf{C}_{3i}$}
\label{sec: C_3i}

The group $\mathbf{C}_{3i}=\mathbf{C}_{3}\times\mathbf{C}_i$ is generated by the threefold rotations $C_{3z}$ about the vertical ($z$) axis and by inversion $I$. 
The transformation of the triplet gap function under inversion is fixed by the condition $\bmd(-\bk)=-\bmd(\bk)$, so we focus on the response of $\bmd$ to the rotations. According to Table \ref{table: Gamma-point-coreps}, 
there are two double-valued coreps at the $\Gamma$ point: $(\Gamma_4,\Gamma_5)$, which is equivalent to the spin-$1/2$ corep, and $\Gamma_6$, which is not. 
As mentioned above, the corep parity is not important in the single-band case.

\underline{$(\Gamma_4,\Gamma_5)$ bands}. 
In the pseudospin bands, we have $\hat{\cal D}_\Gamma(C_{3z})=\hat{D}^{(1/2)}(C_{3z})$. Therefore, $\hat{\cal R}(C_{3z})=\hat R(C_{3z})$ and the triplet gap function 
transforms under rotations as a vector: 
\begin{equation}
\label{C_3i-d-vector}
  C_{3z}:\ \bmd(\bk)\to C_{3z}\bmd(C^{-1}_{3z}\bk). 
\end{equation}
The group $\mathbf{C}_{3i}$ has six single-valued irreps of either parity, $\Gamma_1^\pm$, $\Gamma_2^\pm$, and $\Gamma_3^\pm$, all 1D (Ref. \onlinecite{Lax-book}). 
To obtain the vector basis functions in the expansion (\ref{d-one-band-expansion}), we use the fact that $\be_3$ corresponds to the $\Gamma_1^+$ irrep, 
while $\be_+=(\be_1+i\be_2)/\sqrt{2}$ and $\be_-=(\be_1-i\be_2)/\sqrt{2}$ correspond to $\Gamma_2^+$ and $\Gamma_3^+$, respectively. The odd scalar basis functions can be chosen in the form
\begin{equation}
\label{C_3i-odd-irreps}
  \varphi_{\Gamma_1^-}=k_z,\quad \varphi_{\Gamma_2^-}=k_+,\quad \varphi_{\Gamma_3^-}=k_-,
\end{equation}
where $k_\pm=k_x\pm ik_y$ (a more general expression for $\varphi_{\Gamma_1^-}$ is given below). Examining all possible product representations $\Gamma_i^+\times\Gamma_j^-$ ($i,j=1,2,3$), we obtain: 
$\bm{\phi}_{\Gamma_1}(\bk)=c_1k_-\be_++c_2k_+\be_-+c_3k_z\be_3$. Since the characters of $\Gamma_1$ are real, one can choose real basis function\cite{LL-3} with $c_1=c_2^*=b$ and $c_3^*=c_3=a$. 
In the same way, we obtain: $\bm{\phi}_{\Gamma_2}(\bk)=b_1k_z\be_++b_2k_-\be_-+b_3k_+\be_3$, where $b_{1,2,3}$ are complex constants. 
We note that the irreps $\Gamma_2$ and $\Gamma_3$ are complex conjugate and can therefore be combined into a single physically irreducible 2D representation $(\Gamma_2,\Gamma_3)$. 
The basis functions of this 2D representation can be chosen in the form $\bm{\phi}_{\Gamma_3}(\bk)=\bm{\phi}_{\Gamma_2}^*(\bk)$. The results are shown in the second column of Table \ref{table: d-vector-C_3i}. 

\underline{$\Gamma_6$ bands}. 
According to Table \ref{table: non-pseudospin-corep-matrices}, the $\Gamma$-point corep matrix is given by $\hat{\cal D}_\Gamma(C_{3z})=-\sigma_0$, therefore 
$\hat{\cal R}(C_{3z})=\mathbb{1}$. Inserting this in Eq. (\ref{singlet-triplet-g-one-band}), we have
\begin{equation}
\label{C_3i-d-not-vector}
  C_{3z}:\ \bmd(\bk)\to \bmd(C^{-1}_{3z}\bk), 
\end{equation}
which means that the triplet gap function does not transform like a vector under rotations. In contrast to Eq. (\ref{C_3i-d-vector}), each component of $\bmd$ independently transforms like a scalar, 
with $\be_1$, $\be_2$, and $\be_3$ all corresponding to the identity irrep $\Gamma_1^+$. From the products $\Gamma_1^+\times\Gamma_j^-=\Gamma_j^-$ ($j=1,2,3$) and Eq. (\ref{C_3i-odd-irreps}), we obtain a real basis function
$\bm{\phi}_{\Gamma_1}(\bk)=a_1k_z\be_1+a_2k_z\be_2+a_3k_z\be_3$. Similarly, since $\Gamma_2$ and $\Gamma_3$ are complex conjugate, we have 
$\bm{\phi}_{\Gamma_2}(\bk)=b_1k_+\be_1+b_2k_+\be_2+b_3k_+\be_3$ and $\bm{\phi}_{\Gamma_3}(\bk)=\bm{\phi}_{\Gamma_2}^*(\bk)$. These results are collected in the third column of
Table \ref{table: d-vector-C_3i}. 

Note that the expression for $\bm{\phi}_{\Gamma_1}(\bk)$ in the $\Gamma_6$ bands shown in Table \ref{table: d-vector-C_3i} implies the presence of a line node in the superconducting gap at $k_z=0$. 
It is easy to see that this line node is accidental, because the general scalar basis function of $\Gamma_1^-$ does not vanish anywhere (except $\bk=\bm{0}$), for instance, one could use
$$
  \varphi_{\Gamma_1^-}=k_z+bk_+^3+b^*k_-^3,
$$
where $b$ is a complex constant. In contrast, the scalar basis functions of $\Gamma_2^-$ and $\Gamma_3^-$ necessarily vanish along the line $k_x=k_y=0$, see Appendix \ref{app: zeros of phis}, therefore the triplet
gap corresponding to the $(\Gamma_2,\Gamma_3)$ representation in the $\Gamma_6$ bands has point nodes on the threefold symmetry axis. 
This is different from the $(\Gamma_4,\Gamma_5)$ bands, where the gap functions do not have any symmetry-imposed zeros.

\begin{table}
\caption{The triplet basis functions $\bm{\phi}(\bk)$ for $\mathbb{G}=\mathbf{C}_{3i}$ ($a,a_{1,2,3}$ are real constants, $b,b_{1,2,3}$ are complex constants). 
The single-valued irreps $\gamma$ of $\mathbf{C}_{3i}$ are listed in the first column. Second column: $\bm{\phi}(\bk)$ in the bands in which $\bmd$ transforms like a vector, see Eq. (\ref{C_3i-d-vector}). 
Third column: $\bm{\phi}(\bk)$ in the bands in which $\bmd$ does not transform like a vector, see Eq. (\ref{C_3i-d-not-vector}).}
\begin{tabular}{|c|c|c|}
    \hline
    $\gamma$ & $(\Gamma_4,\Gamma_5)$ bands & $\Gamma_6$ bands  \\ \hline
    $\Gamma_1$   &  $ak_z\be_3+bk_-\be_++b^*k_+\be_-$  &  $a_1k_z\be_1+a_2k_z\be_2+a_3k_z\be_3$  \\ \hline
    \ $(\Gamma_2,\Gamma_3)$\ \   & \ $b_1k_z\be_++b_2k_-\be_-+b_3k_+\be_3$,\ \  &  \ $b_1k_+\be_1+b_2k_+\be_2+b_3k_+\be_3$,\ \   \\ 
                            &  $b_2^*k_+\be_++b_1^*k_z\be_-+b_3^*k_-\be_3$ &  $b_1^*k_-\be_1+b_2^*k_-\be_2+b_3^*k_-\be_3$ \\ \hline
\end{tabular}
\label{table: d-vector-C_3i}
\end{table}

\subsection{$\mathbb{G}=\mathbf{D}_{3d}$}
\label{sec: D_3d}

The group $\mathbf{D}_{3d}=\mathbf{D}_{3}\times\mathbf{C}_i$ is generated by $C_{3z}$ and the twofold rotations $C_{2y}$ about the horizontal ($y$) axis, and also by $I$.
It describes the symmetry of the topological insulator Bi$_2$Se$_3$, which becomes superconducting upon doping.\cite{SC-Bi2Se3} 
There are two double-valued coreps at the $\Gamma$ point: the pseudospin corep $\Gamma_4$ and the non-pseudospin corep $(\Gamma_5,\Gamma_6)$.

\underline{$\Gamma_4$ bands}. 
In this case, we have $\hat{\cal R}(g)=\hat R(g)$, therefore $\bmd$ transforms under rotations as a vector, see Eq. (\ref{d-transform-pseudospin}). The group $\mathbf{D}_{3d}$ has six single-valued irreps, all real: $\Gamma_1^\pm$ and $\Gamma_2^\pm$, which are 1D, and $\Gamma_3^\pm$, 
which is 2D. Observing that $\be_{1,2}$ (or $\be_{\pm}$) transform according to $\Gamma_3^+$ and $\be_3$ transforms according to $\Gamma_2^+$, and using the odd scalar basis functions
\begin{equation}
\label{D_3d-odd-irreps}
  \varphi_{\Gamma_1^-}=i(k_+^3-k_-^3),\quad \varphi_{\Gamma_2^-}=k_z,\quad (\varphi_{\Gamma_3^-,1},\varphi_{\Gamma_3^-,2})=(k_+,k_-),
\end{equation}
we arrive at the expressions in the second column of Table \ref{table: d-vector-D_3d}. Note that $\varphi_{\Gamma_2^-}$ does not have to vanish in the $k_z=0$ plane, see Appendix \ref{app: zeros of phis}, 
and can be written in a more general form as follows:
$$
  \varphi_{\Gamma_2^-}=k_z+a(k_+^3+k_-^3),
$$
where $a$ is a real constant. The zeros of the other basis functions in Eq. (\ref{D_3d-odd-irreps}) are required by symmetry. Therefore, the triplet gap has point nodes at $k_x=k_y=0$ in the $\Gamma_2$ pairing channel, 
and no symmetry-imposed nodes in the $\Gamma_1$ and $\Gamma_3$ channels.

\underline{$(\Gamma_5,\Gamma_6)$ bands}. 
From Table \ref{table: non-pseudospin-corep-matrices} we obtain: $\hat{\cal R}(C_{3z})=\mathbb{1}$ and $\hat{\cal R}(C_{2y})=\hat R(C_{2y})$, \textit{i.e.}, the triplet gap function does not transform like a vector:
\begin{equation}
\label{D_3d-d-not-vector}
  C_{3z}:\ \bmd(\bk)\to\bmd(C^{-1}_{3z}\bk),\quad C_{2y}:\ \bmd(\bk)\to C_{2y}\bmd(C_{2y}\bk). 
\end{equation}
Therefore, $\be_1$ and $\be_3$ correspond to $\Gamma_2^+$, while $\be_2$ corresponds to $\Gamma_1^+$. Using Eq. (\ref{D_3d-odd-irreps}), we obtain the expressions in the third column of Table \ref{table: d-vector-D_3d}. 
The gap function has point nodes at $k_x=k_y=0$ in the $\Gamma_3$ channel, and no symmetry-imposed nodes in the $\Gamma_1$ and $\Gamma_2$ channels.

\begin{table}
\caption{The triplet basis functions $\bm{\phi}(\bk)$ for $\mathbb{G}=\mathbf{D}_{3d}$ ($a_{1,2,3}$ are real constants). 
The single-valued irreps $\gamma$ of $\mathbf{D}_{3d}$ are listed in the first column. Second column: $\bm{\phi}(\bk)$ in the bands in which $\bmd$ transforms like a vector. 
Third column: $\bm{\phi}(\bk)$ in the bands in which $\bmd$ does not transform like a vector, see Eq. (\ref{D_3d-d-not-vector}).}
\begin{tabular}{|c|c|c|}
    \hline
    $\gamma$ & $\Gamma_4$ bands & $(\Gamma_5,\Gamma_6)$ bands  \\ \hline
    \ $\Gamma_1$\ \  &  $a_1(k_x\be_1+k_y\be_2)+a_2k_z\be_3$  &  $a_1k_z\be_1+ia_2(k_+^3-k_-^3)\be_2+a_3k_z\be_3$ \\ \hline
    $\Gamma_2$   &  \ $a_1(k_y\be_1-k_x\be_2)+ia_2(k_+^3-k_-^3)\be_3$\ \  &  \ $ia_1(k_+^3-k_-^3)\be_1+a_2k_z\be_2+ia_3(k_+^3-k_-^3)\be_3$\ \  \\ \hline
    $\Gamma_3$   &  $a_1k_z\be_++a_2k_-\be_-+a_3k_+\be_3$, &  $a_1k_+\be_1+a_2k_+\be_2+a_3k_+\be_3$, \\ 
                 &  $a_2k_+\be_++a_1k_z\be_-+a_3k_-\be_3$ &  $a_1k_-\be_1-a_2k_-\be_2+a_3k_-\be_3$\\ \hline
\end{tabular}
\label{table: d-vector-D_3d}
\end{table}

\subsection{$\mathbb{G}=\mathbf{C}_{6h}$}
\label{sec: C_6h}

The group $\mathbf{C}_{6h}=\mathbf{C}_{6}\times\mathbf{C}_i$ is generated by the sixfold rotations $C_{6z}$ and by $I$. 
There are three double-valued coreps at the $\Gamma$ point: $(\Gamma_7,\Gamma_8)$, $(\Gamma_9,\Gamma_{10})$, and $(\Gamma_{11},\Gamma_{12})$.

\underline{$(\Gamma_7,\Gamma_8)$ and $(\Gamma_9,\Gamma_{10})$ bands}. 
It follows from Table \ref{table: non-pseudospin-corep-matrices} that $\hat{\cal R}(C_{6z})=\hat R(C_{6z})$ in both 
$(\Gamma_7,\Gamma_8)$ and $(\Gamma_9,\Gamma_{10})$ bands, despite the fact that the latter are non-pseudospin ones. Therefore, in these cases $\bmd$ transforms under rotations as a vector, 
see Eq. (\ref{d-transform-pseudospin}). The group $\mathbf{C}_{6h}$ has twelve single-valued irreps $\Gamma_i^\pm$ ($i=1,...,6$), which are all 1D, with $\be_+$, $\be_-$, and $\be_3$ corresponding
to $\Gamma_5^+$, $\Gamma_6^+$, and $\Gamma_1^+$, respectively. Using the odd scalar basis functions 
\begin{equation}
\label{C_6h-odd-irreps}
  \varphi_{\Gamma_1^-}=k_z,\quad \varphi_{\Gamma_2^-}=k_-^2k_z,\quad \varphi_{\Gamma_3^-}=k_+^2k_z,\quad 
  \varphi_{\Gamma_4^-}=bk_+^3+b^*k_-^3,\quad \varphi_{\Gamma_5^-}=k_+,\quad \varphi_{\Gamma_6^-}=k_-,
\end{equation}
where $b$ is a complex constant, we obtain the expressions in the second column of Table \ref{table: d-vector-C_6h}. 
According to Appendix \ref{app: zeros of phis}, the zeros of the basis functions (\ref{C_6h-odd-irreps}) are all required by symmetry, therefore $\bmd(\bk)$ has point nodes at $k_x=k_y=0$ 
in the $(\Gamma_2,\Gamma_3)$ and $\Gamma_4$ pairing channels, and no symmetry-imposed nodes in the $\Gamma_1$ and $(\Gamma_5,\Gamma_6)$ channels.
The $\Gamma_1$ and $\Gamma_4$ irreps are real, while $(\Gamma_2,\Gamma_3)$ and $(\Gamma_5,\Gamma_6)$ form complex conjugate pairs and can be combined into 2D physically irreducible representations, 
as discussed in Sec. \ref{sec: C_3i}. 

\underline{$(\Gamma_{11},\Gamma_{12})$ bands}. From Table \ref{table: non-pseudospin-corep-matrices} we obtain: $\hat{\cal R}(C_{6z})=\hat R(C_{2z})$, \textit{i.e.}, the triplet gap function does not transform like a vector:
\begin{equation}
\label{C_6h-d-not-vector}
  C_{6z}:\ \bmd(\bk)\to C_{2z}\bmd(C^{-1}_{6z}\bk). 
\end{equation}
Therefore, $\be_1$ and $\be_2$ correspond to $\Gamma_4^+$, while $\be_3$ corresponds to $\Gamma_1^+$. Examining the product representations and using Eq. (\ref{C_6h-odd-irreps}), we arrive at the 
basis functions in the third column of Table \ref{table: d-vector-C_6h}. The gap functions in the $\Gamma_1$ and $\Gamma_4$ pairing channels have no symmetry-imposed
nodes, while the $(\Gamma_2,\Gamma_3)$ and $(\Gamma_5,\Gamma_6)$ gap functions have point nodes on the sixfold axis.

\begin{table}
\caption{The triplet basis functions $\bm{\phi}(\bk)$ for $\mathbb{G}=\mathbf{C}_{6h}$ ($a,a_{1,2}$ are real constants, $b,b_{1,2,3}$ are complex constants). 
The single-valued irreps $\gamma$ of $\mathbf{C}_{6h}$ are listed in the first column. Second column: $\bm{\phi}(\bk)$ in the bands in which $\bmd$ transforms like a vector. 
Third column: $\bm{\phi}(\bk)$ in the bands in which $\bmd$ does not transform like a vector, see Eq. (\ref{C_6h-d-not-vector}).}
\begin{tabular}{|c|c|c|}
    \hline
    $\gamma$ & $(\Gamma_7,\Gamma_8)$ and $(\Gamma_9,\Gamma_{10})$ bands & $(\Gamma_{11},\Gamma_{12})$ bands  \\ \hline
    $\Gamma_1$   &  $ak_z\be_3+bk_-\be_++b^*k_+\be_-$  &  \ $ak_z\be_3+(b_1k_+^3+b_1^*k_-^3)\be_1+(b_2k_+^3+b_2^*k_-^3)\be_2$\ \   \\ \hline
    \ $(\Gamma_2,\Gamma_3)$\ \  &  $ak_-^2k_z\be_3+(b_1k_+^3+b_2k_-^3)\be_++b_3k_-\be_-$,  &  $ak_-^2k_z\be_3+b_1k_+\be_1+b_2k_+\be_2$,  \\ 
                            &  $ak_+^2k_z\be_3+b_3^*k_+\be_++(b_1^*k_-^3+b_2^*k_+^3)\be_-$  &  $ak_+^2k_z\be_3+b_1^*k_-\be_1+b_2^*k_-\be_2$  \\ \hline
    $\Gamma_4$   & \ $b_1k_+^2k_z\be_++b_1^*k_-^2k_z\be_-+(b_2k_+^3+b_2^*k_-^3)\be_3$\ \   &  $a_1k_z\be_1+a_2k_z\be_2+(bk_+^3+b^*k_-^3)\be_3$  \\ \hline
    $(\Gamma_5,\Gamma_6)$   &  $ak_+\be_3+b_1k_z\be_++b_2k_+^2k_z\be_-$,  &  $ak_+\be_3+b_1k_-^2k_z\be_1+b_2k_-^2k_z\be_2$, \\ 
                            &  $ak_-\be_3+b_2^*k_-^2k_z\be_++b_1^*k_z\be_-$  &  $ak_-\be_3+b_1^*k_+^2k_z\be_1+b_2^*k_+^2k_z\be_2$ \\ \hline
\end{tabular}
\label{table: d-vector-C_6h}
\end{table}

\subsection{$\mathbb{G}=\mathbf{D}_{6h}$}
\label{sec: Gamma-9}

The group $\mathbf{D}_{6h}=\mathbf{D}_{6}\times\mathbf{C}_i$ is generated by the rotations $C_{6z}$ and $C_{2y}$, and by $I$. This group describes, for instance, the symmetry of the popular heavy-fermion superconductor 
UPt$_3$ (Ref. \onlinecite{UPt3-review}). There are three double-valued coreps at the $\Gamma$ point, corresponding to the pseudospin bands $\Gamma_7$ and the non-pseudospin bands $\Gamma_8$ and $\Gamma_9$.

\underline{$\Gamma_7$ and $\Gamma_8$ bands}.
From Table \ref{table: non-pseudospin-corep-matrices} we obtain that $\hat{\cal R}(g)=\hat R(g)$, therefore $\bmd$
transforms under rotations as a vector, see Eq. (\ref{d-transform-pseudospin}). The group $\mathbf{D}_{6h}$ has twelve single-valued irreps, all real: $\Gamma_{1,2,3,4}^\pm$, which are 1D, and $\Gamma_{5,6}^\pm$,  
which are 2D. The vectors $\be_+$ and $\be_-$ transform according to $\Gamma_5^+$, while $\be_3$ transforms according to $\Gamma_2^+$. Using the odd scalar basis functions 
\begin{eqnarray}
\label{D_6h-odd-irreps}
  && \varphi_{\Gamma_1^-}=i(k_+^6-k_-^6)k_z,\quad \varphi_{\Gamma_2^-}=k_z,\quad \varphi_{\Gamma_3^-}=i(k_+^3-k_-^3),\quad \varphi_{\Gamma_4^-}=k_+^3+k_-^3,\nonumber\\
  && \\
  && (\varphi_{\Gamma_5^-,1},\varphi_{\Gamma_5^-,2})=(k_+,k_-),\quad (\varphi_{\Gamma_6^-,1},\varphi_{\Gamma_6^-,2})=(k_+^2k_z,k_-^2k_z),\nonumber
\end{eqnarray}
we obtain the expressions in the second column of Table \ref{table: d-vector-D_6h}. According to Appendix \ref{app: zeros of phis}, the zeros of the basis functions (\ref{D_6h-odd-irreps}) 
are all required by symmetry, therefore the triplet gap has point nodes on the sixfold axis in the $\Gamma_2$, $\Gamma_3$, $\Gamma_4$, and $\Gamma_6$ pairing channels, and no symmetry-imposed nodes 
in the $\Gamma_1$ and $\Gamma_5$ channels.

\underline{$\Gamma_9$ bands}.
It follows from Table \ref{table: non-pseudospin-corep-matrices} that $\hat{\cal R}(C_{6z})=\hat R(C_{2z})$ and $\hat{\cal R}(C_{2y})=\hat R(C_{2y})$, \textit{i.e.}, the triplet gap function does not transform like a vector:
\begin{equation}
\label{D_6h-d-not-vector}
  C_{6z}:\ \bmd(\bk)\to C_{2z}\bmd(C^{-1}_{6z}\bk),\quad C_{2y}:\ \bmd(\bk)\to C_{2y}\bmd(C_{2y}\bk). 
\end{equation}
Therefore, $\be_1$ corresponds to $\Gamma_4^+$, $\be_2$ to $\Gamma_3^+$, and $\be_3$ to $\Gamma_2^+$. Using Eq. (\ref{D_6h-odd-irreps}), we obtain the expressions in the third column of Table \ref{table: d-vector-D_6h}.
The gap function has no symmetry-imposed nodes in the $\Gamma_1$, $\Gamma_3$, and $\Gamma_4$ pairing channels, and point nodes on the sixfold axis in the $\Gamma_2$, $\Gamma_5$, and $\Gamma_6$ channels. 
The fact that the triplet pairing in the $\Gamma_9$ bands does not have the form predicted by the standard pseudospin-based approach has been previously noticed in Ref. \onlinecite{NHI16}.

\begin{table}
\caption{The triplet basis functions $\bm{\phi}(\bk)$ for $\mathbb{G}=\mathbf{D}_{6h}$ ($a_{1,2,3}$ are real constants). 
The single-valued irreps $\gamma$ of $\mathbf{D}_{6h}$ are listed in the first column. Second column: $\bm{\phi}(\bk)$ in the bands in which $\bmd$ transforms like a vector. 
Third column: $\bm{\phi}(\bk)$ in the bands in which $\bmd$ does not transform like a vector, see Eq. (\ref{D_6h-d-not-vector}).}
\begin{tabular}{|c|c|c|}
    \hline
    $\gamma$ & $\Gamma_7$ and $\Gamma_8$ bands & $\Gamma_9$ bands  \\ \hline
    $\Gamma_1$   &  $a_1(k_x\be_1+k_y\be_2)+a_2k_z\be_3$  &  $a_1(k_+^3+k_-^3)\be_1+ia_2(k_+^3-k_-^3)\be_2+a_3k_z\be_3$  \\ \hline
    $\Gamma_2$   &  $a_1(k_y\be_1-k_x\be_2)+ia_2(k_+^6-k_-^6)k_z\be_3$  & \ $ia_1(k_+^3-k_-^3)\be_1+a_2(k_+^3+k_-^3)\be_2+ia_3(k_+^6-k_-^6)k_z\be_3$\ \   \\ \hline
    $\Gamma_3$   &  $a_1(k_+^2k_z\be_++k_-^2k_z\be_-)+a_2(k_+^3+k_-^3)\be_3$  &  $a_1k_z\be_1+ia_2(k_+^6-k_-^6)k_z\be_2+a_3(k_+^3+k_-^3)\be_3$  \\ \hline
    $\Gamma_4$   &  \ $ia_1(k_+^2k_z\be_+-k_-^2k_z\be_-)+ia_2(k_+^3-k_-^3)\be_3$\ \   &  $ia_1(k_+^6-k_-^6)k_z\be_1+a_2k_z\be_2+ia_3(k_+^3-k_-^3)\be_3$  \\ \hline
    $\Gamma_5$   &  $a_1k_z\be_++a_2k_+\be_3$, $a_1k_z\be_-+a_2k_-\be_3$  &   $a_1k_-^2k_z\be_1+a_2k_-^2k_z\be_2+a_3k_+\be_3$,  \\ 
                 &    &   $a_1k_+^2k_z\be_1-a_2k_+^2k_z\be_2+a_3k_-\be_3$ \\ \hline
    \ $\Gamma_6$\ \    &  $a_1k_+\be_++a_2k_+^2k_z\be_3$, $a_1k_-\be_-+a_2k_-^2k_z\be_3$ &  $a_1k_+\be_1+a_2k_+\be_2+a_3k_-^2k_z\be_3$, \\
                 &    & $a_1k_-\be_1-a_2k_-\be_2+a_3k_+^2k_z\be_3$ \\ \hline
\end{tabular}
\label{table: d-vector-D_6h}
\end{table}

\subsection{Summary}
\label{sec: summary}

The structure of triplet superconducting states in the pairing channel corresponding to an odd $d$-dimensional single-valued irreducible 
(or physically irreducible) representation $\gamma$ of $\mathbb{G}$ is determined by the expansion (\ref{d-one-band-expansion}). According to Eq. (\ref{singlet-triplet-g-one-band}), 
the basis functions transform as
\begin{equation}
\label{phi-to-phi-D-gamma}
  g:\ \bm{\phi}_{a}(\bk)\to {\cal R}(g)\bm{\phi}_{a}(g^{-1}\bk)=\sum_{b=1}^d\bm{\phi}_{b}(\bk)D^{(\gamma)}_{ba}(g),
\end{equation}
where $\hat D^{(\gamma)}$ is the representation matrix (in all cases considered in this paper, $d=1$ or $2$). It follows from Eqs. (\ref{d-one-band-expansion}) and (\ref{phi-to-phi-D-gamma}) 
that the symmetry properties of the order parameter components $\eta_a$ are not affected by the form of ${\cal R}(g)$. Therefore, the Ginzburg-Landau free energy expansion in powers of $\eta_a$ 
depends on the irrep $\gamma$, but not on the character of the electron band states.  

Each basis function can be represented as a linear combination of the products $\be_i\varphi_\mu(\bk)$, where $\be_{1,2,3}$ transform under the point group operations as 
$$
  g:\ \be_i\to {\cal R}(g)\be_i=\sum_j\be_j{\cal R}_{ji}(g)
$$
and $\varphi_\mu(\bk)$ are the scalar basis functions of the odd irreps of $\mathbb{G}$. 
In all pseudospin bands and some non-pseudospin bands, $\hat{\cal R}(g)$ is equal to the rotation matrix $\hat R$ for all proper 
($g=R$) and improper ($g=IR$) rotations, therefore $\be_i$ and $\bm{\phi}$ transform as pseudovectors. 
However, in certain non-pseudospin bands, see Eq. (\ref{exceptional-coreps}), we have $\hat{\cal R}\neq\hat R$ for some $g$, therefore $\be_i$ and $\bm{\phi}$ do not transform as pseudovectors.
This leads to significant changes in the momentum dependence of the triplet gap, see the last columns in Tables \ref{table: d-vector-C_3i}, \ref{table: d-vector-D_3d}, \ref{table: d-vector-C_6h}, 
and \ref{table: d-vector-D_6h}. Note that $\bmd(\bk)$ can have zeros imposed by symmetry only along the main symmetry axis, so that Blount's theorem about the absence of line nodes in the triplet states\cite{Blount85} 
holds in the non-pseudospin bands as well. 

According to Eq. (\ref{singlet-triplet-transform-K}), under TR we have $\bmd(\bk)\to\bmd^*(\bk)$. Since the basis functions for all 1D irreps can be chosen real, the action of TR on a one-component order parameter 
is equivalent to complex conjugation, $\eta\to\eta^*$. This is not automatically the case for the two-component order parameters, since the basis functions of the 2D representations in Tables \ref{table: d-vector-C_3i}, 
\ref{table: d-vector-D_3d}, \ref{table: d-vector-C_6h}, and \ref{table: d-vector-D_6h} have been chosen to satisfy $\bm{\phi}_{2}(\bk)=\bm{\phi}_{1}^*(\bk)$. However, one can make them real by a unitary transformation:
$\tilde{\bm{\phi}}_{1}=(\bm{\phi}_{1}+\bm{\phi}_{2})/\sqrt{2}$ and $\tilde{\bm{\phi}}_{2}=(\bm{\phi}_{1}-\bm{\phi}_{2})/\sqrt{2}i$. Then, the corresponding order parameter transforms under TR into 
its complex conjugate.

One cannot expect that the polynomial expressions for the basis functions in Tables \ref{table: d-vector-C_3i}, \ref{table: d-vector-D_3d}, \ref{table: d-vector-C_6h}, and \ref{table: d-vector-D_6h} 
can correctly describe the momentum dependence of $\bmd(\bk)$ in the systems with a complicated band structure, especially if the Fermi surfaces cross the BZ boundaries. 
In those cases, periodicity of the reciprocal space should be taken into account.

\section{Conclusions}
\label{sec: Conclusion}

The commonly used assumption that the electron Bloch states in the presence of SOC can be chosen to transform under the point group operations in the same way as the pure spin states is not always applicable. 
Specifically, it fails in the bands which correspond to non-pseudospin corepresentations of the magnetic point group of the crystal at the $\Gamma$ point. In general, 
one can define the Bloch bases in twofold degenerate bands across the BZ using a generalization of the Ueda-Rice formula, which satisfies the symmetry and continuity requirements. 

Non-pseudospin character of the Bloch states has profound effects on the symmetry of superconducting pairing. We have shown that in a general multiband case the singlet and triplet gap functions do not transform under 
the point group operations as scalars and pseudovectors, respectively, with the transformation properties of the interband pairing being particularly messy. Leaving investigation of the rich consequences of this observation
to a future work, in this paper we focused on the single-band case and showed that the triplet gap function $\bmd(\bk)$ does not transform like a pseudovector in four exceptional cases, 
corresponding to non-pseudospin bands in trigonal and hexagonal crystals, see Eq. (\ref{exceptional-coreps}). 

The momentum dependence of the triplet gaps in the exceptional cases differs considerably from that obtained using the standard pseudospin-based approach.\cite{SU-review,TheBook} 
In particular, all two-component non-pseudospin gap functions have point nodes along the main symmetry axis, while some of the point nodes of the one-component gap functions in the pseudospin bands are absent 
in the non-pseudospin ones. In agreement with Blount's theorem, there are no symmetry-protected line nodes in the gap. Our predictions about the nodal structure can be tested in experiment, 
although their direct application may be hindered by a complicated band structure and by the lack of knowledge about the $\Gamma$-point corepresentation realized in a given superconducting band.

\acknowledgments

This work was supported by a Discovery Grant 2015-06656 from the Natural Sciences and Engineering Research Council of Canada.

\appendix

\section{Non-pseudospin coreps for $\mathbf{D}_{3d}$}
\label{app: D_3d-coreps}

In this appendix, we show how to obtain the $\Gamma$-point corep matrices and the basis functions in Table \ref{table: non-pseudospin-corep-matrices}, 
using as an example a trigonal crystal with $\mathbb{G}=\mathbf{D}_{3d}$. 
The corresponding group $\tilde{\mathbb{G}}=\mathbf{D}_{3}$ is generated by the rotations $C_{3z}$ and $C_{2y}$ and has three double-valued irreps: $\Gamma_4$, which is 2D, and also $\Gamma_5$ and $\Gamma_6$, which are 1D 
and complex conjugate to each other.\cite{Lax-book,BC-book} The $\Gamma_4$ irrep is equivalent to the spin-$1/2$ irrep. 

The irreps $\Gamma_5$ and $\Gamma_6$ pair up to form a single ``physically irreducible'' 2D representation of $\mathbf{D}_{3}$, or a 2D corep of the magnetic point group $\mathbf{D}_{3}+{\cal C}\mathbf{D}_{3}$, 
which is given by the following matrices:
$$
  \hat{\cal D}(g)=\left(\begin{array}{cc}
              \chi_{\Gamma_5}(g) & 0 \\
              0 & \chi_{\Gamma_6}(g)
              \end{array}\right),\quad 
  \hat{\cal D}({\cal C})=\left(\begin{array}{cc}
              0 & -1 \\
              1 & 0
              \end{array}\right),
$$
where $\chi_{\Gamma_5}$ and $\chi_{\Gamma_6}=\chi^*_{\Gamma_5}$ are the group characters of $\mathbf{D}_{3}$. This corep is classified as being of ``pairing'' type (Ref. \onlinecite{Lax-book}) or 
``Case C'' (Ref. \onlinecite{BC-book}). From the character tables, we obtain: 
\begin{equation}
\label{app: Gamma_56-corep-1}
    \hat{\cal D}(C_{3z})=\left(\begin{array}{cc}
              -1 & 0 \\
              0 & -1
              \end{array}\right),\quad
    \hat{\cal D}(C_{2y})=\left(\begin{array}{cc}
              -i & 0 \\
              0 & i
              \end{array}\right).
\end{equation}
Comparing these matrices with $D^{(1/2)}(C_{3z})$ and $D^{(1/2)}(C_{2y})$, respectively, we see that the $(\Gamma_5,\Gamma_6)$ corep is not equivalent to the spin-$1/2$ representation. Therefore, the corresponding
Bloch states at the $\Gamma$ point, $|\bm{0},1\rangle$ and $|\bm{0},2\rangle={\cal C}|\bm{0},1\rangle$, do not transform under the point group rotations as the basis spinors $\xi_1$ and $\xi_2$. 

It is easy to show that the basis of the $(\Gamma_5,\Gamma_6)$ corep which reproduces the matrices (\ref{app: Gamma_56-corep-1}) can be formally written as $(\phi_1,\phi_2)\propto(\xi_1^3+i\xi_2^3,\xi_2^3+i\xi_1^3)$. 
This corep can be brought by a unitary transformation to an equivalent form
\begin{equation}
\label{app: Gamma_56-corep-2}
    \hat{\cal D}(C_{3z})=\left(\begin{array}{cc}
              -1 & 0 \\
              0 & -1
              \end{array}\right),\quad
    \hat{\cal D}(C_{2y})=\left(\begin{array}{cc}
              0 & -1 \\
              1 & 0
              \end{array}\right),
\end{equation}
which corresponds to the basis $(\phi_1,\phi_2)\propto(\xi_1^3,\xi_2^3)$. 
The basis functions with an explicit coordinate dependence can be obtained by observing that $\rho_\pm=x\pm iy$ and $\xi_{1,2}$ transform under the elements of the magnetic point group as follows:
\begin{eqnarray*}
  && C_{3z}\rho_\pm=e^{\mp 2i\pi /3}\rho_\pm,\quad C_{2y}\rho_\pm=-r_\mp,\quad {\cal C}(c\rho_\pm)=-c^*r_\mp\\
  && C_{3z}\xi_{1,2}=e^{\mp i\pi/3},\quad C_{2y}\xi_{1,2}=\pm\xi_{2,1},\quad {\cal C}(c\xi_{1,2})=\pm c^*\xi_{2,1},
\end{eqnarray*}
where we included a $c$-number coefficient to emphasize the antilinearity of the conjugation operator. One possible choice of odd basis functions that change sign under $C_{3z}$ is
$\phi_1\propto \rho_+\xi_1$ and $\phi_2\propto \rho_-\xi_2$. Applying $C_{2y}$, see Eq. (\ref{app: Gamma_56-corep-2}), and also imposing the conditions that $\phi_2={\cal C}\phi_1$ produces the following result:
\begin{equation}
\label{app: phi_12}
  \phi_1=a\rho_+\xi_1,\quad \phi_2=-a\rho_-\xi_2,
\end{equation}
where $a$ is a real constant.

The corep basis functions for the uniaxial point groups can be chosen to have a definite value of $\hat j_z=\hat l_z+\hat s_z$ -- the projection of the total angular momentum on the main symmetry axis, where 
$\hat l_z$ is the orbital angular momentum and $\hat s_z=\hat\sigma_3/2$ is the spin angular momentum. In particular, for the expressions (\ref{app: phi_12}) we have
$\hat j_z\phi_1=(3/2)\phi_1$ and $\hat j_z\phi_2=(-3/2)\phi_2$. The basis functions shown in Table \ref{table: non-pseudospin-corep-matrices} correspond to the lowest possible values of $|j_z|$ 
compatible with the symmetry requirements.

\section{Zeros of $\varphi(\bk)$}
\label{app: zeros of phis}

Let us consider an odd $r$-dimensional single-valued irrep $\gamma^-$ of $\mathbb{G}$, with the scalar basis functions $\varphi_\mu(\bk)$, where $\mu=1,...,r$. 
For the four point groups $\mathbf{C}_{3i}$, $\mathbf{D}_{3d}$, $\mathbf{C}_{6h}$, and $\mathbf{D}_{6h}$, see Eq. (\ref{exceptional-coreps}), we have $r=1$ or $2$. 
One can show that, apart from the trivial zero at $\bk=\bm{0}$, the basis functions for some irreps vanish along the line $k_x=k_y=0$ and/or in the plane $k_z=0$.  

The action of the rotations $C_{nz}$ ($n=2,3,6$) and $C_{2y}$ on the basis functions is given by $C_{nz}\varphi_\mu(\bk)=\varphi_\mu(C_{nz}^{-1}\bk)$ and $C_{2y}\varphi_\mu(\bk)=\varphi_\mu(C_{2y}\bk)$. In particular,
$C_{nz}\varphi_\mu(0,0,k_z)=\varphi_\mu(0,0,k_z)$ and $C_{2y}\varphi_\mu(0,0,k_z)=\varphi_\mu(0,0,-k_z)=-\varphi_\mu(0,0,k_z)$.
Therefore, $\varphi_\mu(0,0,k_z)=0$, unless $\chi_{\gamma^-}(C_{nz})=r$ and $\chi_{\gamma^-}(C_{2y})=-r$. Inspecting the group character tables, we obtain that the basis functions 
vanish along the $k_x=k_y=0$ line for the following irreps: $\Gamma_{2,3}$ of $\mathbf{C}_{3i}$, $\Gamma_{1,3}$ of $\mathbf{D}_{3d}$, 
$\Gamma_{2,3,4,5,6}$ of $\mathbf{C}_{6h}$, and $\Gamma_{1,3,4,5,6}$ of $\mathbf{D}_{6h}$.

If $C_{2z}$ is a symmetry element, then its action on the basis functions is given by $C_{2z}\varphi_\mu(\bk)=\varphi_\mu(C_{2z}\bk)$, in particular,
$C_{2z}\varphi_\mu(k_x,k_y,0)=\varphi_\mu(-k_x,-k_y,0)=-\varphi_\mu(k_x,k_y,0)$.
Therefore, $\varphi_\mu(k_x,k_y,0)=0$, unless $\chi_{\gamma^-}(C_{2z})=-r$. Using the group character tables, we obtain that the basis functions 
vanish in the $k_z=0$ plane for the following irreps: $\Gamma_{1,2,3}$ of $\mathbf{C}_{6h}$ and $\Gamma_{1,2,6}$ of $\mathbf{D}_{6h}$.

\end{document}